\documentstyle[prl,aps,multicol]{revtex}

\begin{document}
\renewcommand{\narrowtext}{\begin{multicols}{2}
\global\columnwidth20.5pc} 
\renewcommand{\widetext}{\end{multicols}
\global\columnwidth42.5pc} \multicolsep = 8pt plus 4pt minus 3pt

\title{Dynamic nuclear polarization at the edge of a two-dimensional 
electron gas}
\author{David C. Dixon, Keith R. Wald, Paul L. McEuen}
\address{Department of Physics, University of California at Berkeley, 
and Materials Sciences Division, E. O. Lawrence Berkeley National 
Laboratory, University of California, Berkeley, CA 94720}
\author{M. R. Melloch}
\address{School of Electrical and Computer Engineering, Purdue 
University, West Lafayette, IN  47907}

\date{\today} 
\maketitle

\begin{abstract}
We have used gated GaAs/AlGaAs heterostructures to explore nonlinear 
transport between spin-resolved Landau level (LL) edge states over a 
submicron region of two-dimensional electron gas (2DEG).  The current 
$I$ flowing from one edge state to the other as a function of the 
voltage $V$ between them shows diode-like behavior---a rapid increase in 
$I$ above a well-defined threshold $V_t$ under forward bias, and a 
slower increase in $I$ under reverse bias.  In these measurements, a 
pronounced influence of a current-induced nuclear spin polarization on 
the spin splitting is observed, and supported by a series of NMR 
experiments.  We conclude that the hyperfine interaction plays an 
important role in determining the electronic properties at the edge of a 
2DEG.
\end{abstract}

\narrowtext
\setlength{\parindent}{1em}
\section{INTRODUCTION}
The physics of two-dimensional electron gases (2DEGs) formed at 
GaAs/AlGaAs heterojunctions has become a very popular field in the last 
several years, owing to the 2DEG's many interesting properties, most 
notably the quantum Hall effect (QHE)\cite{QHE}.  When placed in a 
strong perpendicular magnetic field, the electronic energy levels of the 
2DEG congregate into Landau levels (LLs), whose energies are given by:
\begin{equation}
E=(n+\frac{1}{2}) \hbar \omega_c + g \mu_B B S_z+E_{ex} + A \langle I_z 
\rangle S_z
\end{equation}
The first term of Eq. 1 gives the orbital LL splitting, where $n$ is the 
orbital LL index and $\hbar \omega_c$ is the cyclotron energy.  The 
second term lifts the spin degeneracy of each orbital LL through the 
Zeeman interaction for GaAs,  $g \mu_B B \sim 0.016 \hbar \omega_c$, 
with $S_z$ being the electron spin ($\pm \frac{1}{2}$).  The third term 
expresses the effects of exchange, which depends sensitively on 
temperature and on the filling factor $\nu = n_s h/eB$ (the number of 
LLs filled for 2D electron density $n_s$).  Exchange can affect the 
total energy considerably, sometimes by as much as a few meV.  The final 
term involves the influence of nuclear polarization $\langle I_z 
\rangle$ through the contact hyperfine interaction, the effect of which 
is the focus of our paper and is discussed in more detail later.

Due to their high mobility and ease of fabrication, 2DEGs provide 
a useful medium for examining many-body physical effects, such as 
exchange.  Even though the Zeeman energy splitting is only a tiny 
fraction of the orbital LL splitting, exchange effects favor a 
ferromagnetic ground state near $\nu = 1$, increasing the effective spin 
gap.  It has recently been observed that the low-energy excitations of 
such a spin-polarized 2DEG are not single spin flips, but rather 
spatially extended spin-textures (skyrmions), in which electrons 
gradually tilt their spins from the center of the texture outward, with 
the size of the skyrmion set by the competition between exchange and 
Zeeman energies \cite{Sondhi}.  Skyrmions have been detected using 
various techniques \cite{Barrett,Schmeller} in bulk 2DEGs, underscoring 
the importance of treating the 2DEG as an interacting many-body system.
	
It is recognized that the {\it nuclei} of the GaAs crystal can 
affect the electronic properties of the 2DEG as well.  Any nonzero 
nuclear polarization $\langle I_z \rangle$ will create an extra 
effective magnetic field felt by the electrons, producing an Overhauser 
shift in the electron energies that can be detected with electron spin 
resonance absorption \cite{Dobers}.  In turn, a net electron 
polarization produces a Knight shift in the nuclear energies, which can 
be used to measure the spin polarization of the 2DEG \cite{Barrett}.  In 
addition to these energy shifts, the hyperfine interaction allows 
"flip-flop" scattering in GaAs, where an electron "flips" its spin 
simultaneous with the "flop" of a nuclear spin in the opposite 
direction, conserving the net spin of the entire system.
  
Nuclear spin effects in bulk 2DEGs have been well-studied, but in 
this paper we shall be examining these effects at the edge of the 2DEG. 
When $\nu$ is an integer, all occupied LLs are full and the bulk 2DEG is 
incompressible.  At the edge, however, the electron density gradually 
descends from $\nu$ to zero and the LL energies curve upward, due to the 
electrostatic confinement potential.  The intersections of the LLs with 
the Fermi energy $E_F$ near the edge define regions where electrons can 
be added to the 2DEG.  These "edge states" (or "edge channels") are 
spatially separated independent channels, each carrying an identical 
amount of current at equilibrium \cite{Haug}. Self-consistent 
electrostatic screening modifies the edge states, creating wide 
compressible and incompressible stripes at the edge, with a 
corresponding steplike potential profile (Fig. 2(a)) 
\cite{Chang,McEuen,Chklovskii,Zhit1,Hwang}.
	
The complete many-body physics of the edge is not well understood, 
although theories predict that the edge may exhibit many-body phenomena, 
such as spin textures \cite{Karlhede}.  The relative tininess of the 
edge region makes many measurement techniques unfeasible, but electronic 
transport, which necessarily takes place at the edge in the QH regime, 
provides a probe into the nature of these states.  At equilibrium the 
edge states all maintain the same electrochemical potential.  Using 
submicron gates deposited on top of the heterostructure, however, one 
can  selectively backscatter the edge states, induce different 
potentials in different edge states, and measure the resultant 
inter-edge scattering \cite{Wees}.  Scattering between spin-degenerate 
\cite{Alphenaar} and spin-split \cite{Muller} edge states has been 
considered previously for the linear regime, as has non-linear 
scattering between spin-degenerate edge states \cite{Zhit1,Komiyama}.  
In this paper, we report measurements of non-linear transport between 
spin-split edge states, and show that spin-flip relaxation produces a 
nuclear polarization of the Ga and As nuclei.  This polarization can in 
turn drastically affect the electronic energies at the edge of a 2DEG.

In Section II of this paper, we describe the measurement setup and 
the method by which a potential imbalance is created between spin-split 
edge states using submicron gates.  We also describe a simple picture of 
the edge utilizing the "spin diode" model used by Kane {\it et al.} 
\cite{Kane}.  Section III contains our experimental results, which 
display features that are best explained by dynamic nuclear polarization 
(DNP) of the nuclear spins.  We present strong evidence for this 
interpretation with a series of NMR experiments.  We continue in Section 
IV with some observations about the data, and we briefly discuss some 
possible consequences of our results for models of the spin-split edge.  
In Section V we compare our findings with earlier results by our group 
\cite{Wald}, and we conclude in Section VI.

\section{MEASUREMENT METHODOLOGY}

A schematic diagram of the device under consideration is shown in Fig. 
1(a).  Electrons populated up to an electrochemical potential $\mu = 
-eV$ enter the two spin-split edge channels from contact 1.  Gates A and 
B ("AB split-gate") are tuned so that the upper (inner, spin-down) edge 
state is reflected by the gate's potential barrier, but the lower 
(outer, spin-up) channel is transmitted.  After passing through these 
gates, the outer edge channel, still at potential $\mu$, propagates 
along gate A in close proximity to the grounded inner edge channel.  The 
edge channels are not in equilibrium in this region, so there is a net 
scattering of electrons from one channel to the other.  These scattered 
electrons propagate in the inner edge channel to a current amplifier 
(contact 3) and are measured as current $I$.  Unscattered electrons 
remain in the outer edge channel and pass between gates A and C ("AC 
split-gate") into the grounded contact 2 and avoid detection by the 
current amplifier.  The current $I$ measured in this three-terminal 
arrangement therefore solely originates from interedge scattering.

One may notice in Fig. 1 that the outer edge states are shown going 
underneath gates B and C.  This is because these gates are only 
partially depleted, but depleted enough so that the electron density 
beneath the gate is such that only one LL is filled ($\nu \sim 1$), and 
the inner (spin up) edge state is reflected.  The region of 2DEG between 
the split-gates must also reflect the inner edge state, which can be 
accomplished by increasing the voltage on gate A ($V_A$) to partially 
deplete the 2DEG to $\nu \sim 1$ throughout this region.   The reasons 
for using this semi-depletion method are detailed in Section VI.

A schematic electrochemical energy diagram of the 2DEG edge is shown in 
Fig. 2, where the bulk of the sample is to the left and the edge is to 
the right.  A combination of the sample's electrostatic confinement 
potential and the electrons' ability (or inability) to screen this 
potential leads to the slanting stepwise energy profile shown 
\cite{Chklovskii}.  Electrons in the compressible regions can  move 
around to screen the external confinement potential, creating the 
energetically flat regions shown. The electron density within each 
compressible strip falls steadily from left to right.  Between the 
compressible regions, the electron density is fixed at integer filling 
factor, so these incompressible regions cannot screen the confinement 
potential.  It should be noted that this picture does not include 
quantum mechanical electron-electron interactions such as exchange, 
which complicate the picture considerably.  We will discuss this 
complication in Section V.

The energy level diagram in Fig. 2 resembles that of a diode 
\cite{Kane}, with the spin-split edge states playing the role of the 
diode's $p$- and $n$-doped regions.  When the outer edge channel is 
forward biased, as shown in Fig. 2(b), the energy difference between the 
partially filled states of the inner edge channel and the available 
empty states of the outer edge channel decreases, and the incompressible 
strip between the edge channels becomes narrower \cite{Zhit2}.  For 
small forward bias, only a small current of thermal electrons will flow 
between the edge states, resulting in a small $I$.  Once $|e|V$ exceeds 
the LL energy splitting $g \mu_B B$, however, the incompressible strip 
disappears, and a large current of electrons can move freely from the 
inner to outer edge channels.  We therefore expect a threshold voltage 
$V_t$ in the $I-V$ trace, corresponding to the LL energy splitting.  
Conversely, for negative bias (Fig. 2(c)), the interedge energy 
splitting becomes enhanced, and in order to scatter between edge states, 
electrons must tunnel through the incompressible strip, leading to a 
small $I$ which depends on both the bias $V$ and the width of the tunnel 
barrier (which is itself a function of $V$).  Because of the different 
modes of transport for forward and negative bias, there should be an 
asymmetry in $I$.  Previous experiments on transport between large 
compressible regions \cite{Kane,Zhit2,Kane2} and between spin-degenerate 
edge channels and large compressible regions \cite{Zhit3} have shown 
this asymmetry.

Since the LLs in the spin diode are of opposite spin, the scattering of 
an electron from one LL to the other must be accompanied by a spin flip.  
It is important to note, however, that for forward bias, electrons do 
not necessarily have to flip their spins in order to register a current 
$I$.  They can be excited from the upper LL of the inner edge channel 
(thermally, or with help from a high bias) into the empty states in the 
upper LL of the outer edge channel, and stay in that channel long enough 
to make it through the AC split-gate and disappear into contact 2.  
However, some of these "hot" electrons in the upper LL relax to the 
lower LL by flipping their spin, which can be caused either by 
spin-orbit scattering \cite{Muller} or by the contact hyperfine 
interaction between the electron and the Ga and As nuclei 
\cite{Vagner}.  We will be concerned with the effects of this 
hyperfine-mediated scattering.

\section{EXPERIMENTAL RESULTS AND INTEPRETATION}

The device was fashioned from a GaAs/AlGaAs heterostructure with a 2DEG 
density $n_s = 2.5 \times 10^{11}$ electrons/cm$^2$ and a mobility of 
$\sim 10^6$ cm$^2$/Vs.  Patterned split-gates of layered Cr and Au were 
evaporated on the surface of the structure, and Ni/Ge/Au contacts were 
annealed to make electrical contact with the 2DEG.  The device is shown 
in Fig. 1(b).  The current measurement setup used a virtual-ground 
preamplifier in a standard DC configuration, with the device mounted in 
a dilution refrigerator and cooled to a base temperature of 30 mK. 

For all the spin diode experiments, the magnetic field was set to 7.0 T 
($\nu = 2$) and the AC and AB split-gates were tuned to transmit only 
the outermost edge state, as shown in Fig. 1(a), so that the measurement 
probes the scattering between $n$ = 0$\uparrow$ and $n$ = 0$\downarrow$ 
Landau levels.  A typical $I-V$ measurement is plotted in Fig. 3, 
showing a rapid increase of current in forward bias with a more gradual 
increase in reverse bias, as predicted by the spin diode model described 
in Section II.  Note that the forward-bias threshold voltage $V_t$, 
where $I$ rapidly changes slope, is comparable to, but greater than, the 
bare spin splitting  $g \mu_B B \sim$ 0.18 meV.  This is much less than 
the exchange-enhanced spin splitting (a few meV) in the bulk 2DEG.  We 
will return to this in Section IV.  

We did not observe the complex structure under reverse bias reported by 
Kane {\it et al.} \cite{Kane}, possibly because our device has a 
different geometry than the interrupted Corbino-style device used in 
their experiments.  Also, as we will show in the Discussion section, the 
estimated width of the incompressible region in the Kane spin diodes (70 
nm) is about ten times larger than ours, and as such could be large 
enough to exhibit different many-body effects than what we observe.

An important observation is that the $I-V$ curve in Fig. 3 is 
hysteretic.  The direction of the hysteresis is indicated by the arrows.  
For forward bias, the current is larger sweeping up in bias than when 
sweeping down, and for negative bias, the current is more negative 
sweeping up towards zero bias than when sweeping down away from zero 
bias.  The size of the hysteresis loop depends on the sweep rate; the 
sweep shown in Fig. 3 lasted approximately five minutes.  If the sweep 
is halted at some point in the loop, the current 
exponentially\cite{decay} relaxes to an equilibrium value with a long 
relaxation time, typically on the order of 30 seconds.  

To understand the origin of this hysteresis, we first note that the 
equilibration time constant is similar to previously measured nuclear 
relaxation times for Ga and As in quantum wells\cite{Krapf}, indicating 
that the source of the hysteresis is the influence of the GaAs nuclear 
spins upon the 2DEG electron spin energies through the contact hyperfine 
interaction.  The hyperfine Hamiltonian is:
\begin{equation}
A\vec{I}\cdot \vec{S}= \frac{A}{2}(I^+ S^- + I^- S^+) + A I_z S_z
\end{equation}
where $A$ is the hyperfine constant, and $\vec{I}$ and $\vec{S}$ are the 
nuclear and electron spins, respectively.  The first term of Eq. 2, 
consisting of ladder operators, corresponds to the simultaneous 
flip-flop of electron and nuclear spins, and the second term is the 
hyperfine splitting.  

We connect the hysteresis of Fig. 3 to the hyperfine interaction as 
follows.  In our experiments a steady influx of spin-polarized electrons 
enters through the AB split-gate, dynamically polarizing the nuclei in 
the scattering region through flip-flop scattering. The formation of a 
nuclear polarization $\langle I_z \rangle$ in turn affects the electron 
energies through the Zeeman-like term $A\langle I_z \rangle S_z$, which 
acts like an effective magnetic field $B_{eff} = \langle I_z \rangle / g 
\mu_B$ (Overhauser effect). This extra field changes the LL energy 
splitting to $g \mu_B(B + B_{eff})$, which in turn shifts the threshold 
voltage $V_t$.  
Let us consider that the voltage $V$ begins at large negative bias 
(lower left-hand corner of Fig. 3).  Here the current flow is from outer 
to inner edge states, which involves a spin flip of up to down.  This 
spin flip, through the hyperfine interaction, "flops" a nucleus from 
"down" to "up"\cite{spinflip}, so a steady current flow results in a net 
spin-up nuclear polarization (positive $\langle I_z \rangle$).  When $V$ 
is swept up to positive values, the spin diode is in forward bias, so 
that a large current will begin to flow from inner to outer edge states 
once $V$ reaches the threshold voltage $V_t$.  This threshold, however, 
is not just the bare spin splitting $g \mu_B B$; $\langle I_z \rangle$ 
is still nonzero because of the slow nuclear polarization decay rate, 
and it creates a negative $B_{eff}$ ($g$ = -0.44).  Therefore, $V_t$ is 
lowered and $I$ is increased, compared to the case of unpolarized 
nuclei.  Continuing the sweep, at large positive bias the current is 
from inner to outer edge states, which can involve a spin flip from down 
to up.  This "flops" a nucleus from "up" to "down," so a steady current 
flow in this case pumps the nuclei towards a net spin-down nuclear 
polarization (negative $\langle I_z \rangle$).  A negative $\langle I_z 
\rangle$ creates a positive $B_{eff}$, which increases $V_t$ and 
decreases $I$.  This accounts for the lower branch of the hysteresis 
loop for forward bias in Fig. 3.  To finish the sweep, $V$ goes back to 
negative values, the current flow pumps the nuclei back to a spin-up 
polarization, and the cycle repeats.

The important point of this model is that the current induces a nuclear 
polarization through the flip-flop term of Eq. 2, and is in turn 
affected by the already-existing nuclear polarization through the Zeeman 
term of Eq. 2.  The complex interplay between the two effects, combined 
with the long relaxation times for Ga and As nuclei, leads to the 
observed hysteresis.  

It would be useful to observe these hyperfine effects independently of 
each other by measuring the $I-V$ profile of the spin diode at a 
constant $\langle I_z \rangle$.  To do this, we performed experiments 
where we held $V$ at a fixed value $V_{dwell}$ for 60 seconds-long 
enough for $\langle I_z \rangle$ to reach equilibrium-then quickly 
ramped $V$ to a voltage, measured $I$ at that voltage, and immediately 
returned to $V_{dwell}$ to reset the nuclear polarization.  This small 
duty cycle procedure, repeated for many values of $V$, keeps the system 
in a state of constant nuclear polarization, while measuring the $I-V$ 
profile at this fixed polarization.  Similar experiments were carried 
out by Kane {\it et al.}\cite{Kane}.

Three examples of these measurements, for $V_{dwell}$ = +1, 0, and -1 
mV, are shown in Fig. 4.  According to the model, these $I-V$'s should 
correspond to an enhancement, no effect, and a decrease in the electron 
spin splitting, respectively.  This is indeed what is observed, seeing 
that $V_t$ is shifted by a significant amount between traces.  For 
$V_{dwell}$  = 0 mV, we believe the nuclei remain unpolarized, and the 
threshold $V_t \sim$ 0.27 mV.  This suggests that $g$ is slightly 
enhanced ($g$* $\sim 1.5g$), yet still much smaller than has been 
measured in bulk 2DEGs\cite{Usher}, where $g*$ can be as large as $20g$.  
We interpret the shift $\Delta V_t$ between dwell plots as being the 
Overhauser shift.  For both $V_{dwell}$ = +1 V and -1 V, $e|\Delta V_t| 
= A \langle I_z \rangle S_z \sim$ 0.10 meV, corresponding to an 
effective Overhauser field of about 4 T.  The maximum Overhauser field 
for GaAs\cite{Krapf} is 5.3 T, so the nuclear spins in the scattering 
region must be highly polarized (about 85\%).

To demonstrate further that $I$ is indeed affected by the state of the 
nuclear spins, we performed a series of nuclear magnetic resonance (NMR) 
experiments with the spin diode.  We mounted a simple one-turn coil next 
to our sample, to which we applied a frequency-tunable AC voltage in 
order to produce an AC magnetic field perpendicular to $B$ 
({\it ie.}, in the plane of the 2DEG).  The spin diode was held at 
forward bias $V_{dwell} > V_t$, polarizing the nuclei in the scattering 
region.  Fig. 5 displays I as a function of coil frequency near the 
$^{75}$As resonance, for three slightly different values of B.  For all 
measurements, the frequency was swept from low to high values.  Each 
trace shows a well-defined peak in current, with the peak shifting to 
higher frequencies for increasing $B$. 

The peaks are due to NMR absorption;  matching the in-plane AC magnetic 
field frequency to the NMR absorption energy for a nuclear species 
partially erases the polarization of that species, decreasing the 
Overhauser shift (and $V_t$) and leading to a sudden increase in 
current.  The peak is located at the expected NMR frequency for 
$^{75}$As, and scales appropriately with B.  Similar behavior was seen 
for the $^{69}$Ga and $^{71}$Ga absorption lines\cite{NMR}.  Kane 
{\it et al.}\cite{Kane} reported similar NMR results in their spin 
diode experiments.

The long exponential tail on the right side of the peaks for $B$ = 7.05 
and 7.1 T is due to the long equilibration time, which was comparable to 
the frequency sweeping rate in these measurements.  The $B$ = 7.0 T peak 
was swept much more slowly, so that the nuclei were always close to 
equilibrium during the sweep, as evidenced by the disappearance of the 
long tail.  When the AC frequency is swept very slowly, the widths of 
the NMR features are approximately 20 KHz.  This is on the order of the 
Knight shift expected for the electron density of our 2DEG\cite{Tycko}, 
and we will discuss this further in the next Section. 

We carried out a series of similar diode-like experiments at $\nu = 4$, 
measuring scattering between spin-degenerate orbital LL edge states.  In 
those experiments, we observed asymmetric $I-V$ curves with a threshold 
voltage $V_t$ comparable to the cyclotron energy $\omega_c = eB/m$*.  
More details about these experiments are published 
elsewhere\cite{Frenchpaper}.

\section{DISCUSSION}

We first note that, although our simple model of the spin-split edge 
explains the electron transport data rather well, it does not include 
the well-documented effects of exchange, which have been 
observed\cite{Usher} to greatly increase the spin gap in bulk 2DEGs near 
$\nu = 1$.  These effects have been predicted to manifest themselves at 
the edge as well, particularly in the neighborhood of the $\nu = 1$ 
incompressible strip.  One theory of the spin-split edge\cite{Rijkels} 
predicts that the spin gap in this region can be enhanced by as much as 
a factor of 50.  Our measurements of this gap (through the threshold 
voltage $V_t$) appear to indicate otherwise-the spin gap is only 
slightly enhanced ($g$*$\sim 1.5g$)-but this conclusion is based upon 
the assumption that $V_t$ and  $g$* $\mu_B B$ are directly related.

To estimate the various pertinent length scales, we applied the self-
consistent electrostatic model of Chklovskii {\it et al.} 
\cite{Chklovskii} to spin-split edge states, substituting the bare 
spin gap $g \mu_B B$ for $\hbar\omega_c$.  In this case, the $\nu = 1$ 
incompressible strip is centered at about 70 nm from the edge of the 
2DEG, with a width of about 7 nm, comparable to the magnetic length 
$\lambda$ (10 nm).  At length scales this small, the local density 
approximation fails, so it is reasonable to expect that exchange 
calculations for bulk samples cannot be applied directly to such a small 
edge region.  More sophisticated theories of the physics of spin-split 
edge states do exist, and we discuss their relevance to our experiments 
as follows.

One theory\cite{Rijkels} of spin-split edge states predicts hysteresis 
due entirely to electron-electron interactions.  At a critical potential 
imbalance $\Delta \mu^+_{cr}$, the edge channels are predicted to switch 
positions, remaining in this switched orientation until a different 
potential difference $\Delta \mu^-_{cr}$ is reached.  We believe, 
however, that our DNP interpretation explains the observed hysteresis 
adequately, and we see no compelling evidence of this channel-crossing 
phenomenon.  Another theory\cite{Karlhede} predicts that, for certain 
ranges of the depletion width $w$ (normalized to $\tilde{w} = 
w/\lambda$) and Zeeman strength $\tilde{g} = g \mu_B /(e^2/e\lambda)$, 
the 2DEG edge supports spin deformations running along the edge (for 
$\nu <  1$).  We estimate our device's parameters to be $\tilde{w} \sim 
7$ and $g \sim 0.016$, placing it within the parameter space where these 
spin-textured edges are predicted to exist.  This textured edge theory, 
however, makes no predictions about the transport properties of such a 
system, so we cannot confirm the existence of such a texture in our 
experiment.  We know of no theory which specifically predicts the 
current flow between spin-split edges as a function of the non-linear 
potential difference between them.  Such a theory would require careful 
examination of many different facets of the problem: self-consistent 
electrostatics, exchange interactions, potential imbalances, 
electrodynamic effects due to interedge current flow, and, as we discuss 
below, hyperfine interactions.  

It is clear from the dwell plots in Fig. 4 that a net nuclear 
polarization creates a large Overhauser shift of the edge state 
energies, so we believe that a complete description of the physics of 
the 2DEG edge cannot ignore hyperfine effects.  While it is true that 
edge state transport experiments in the linear regime ({\it ie.} 
$|eV| < g \mu_B B$) will not create a nuclear polarization, it is clear 
from our experiments that non-linear transport between spin-split edges 
can create one, so it is important to consider hyperfine effects in this 
regime.  The many-body effects predicted by theory could very well be 
affected by the nuclear polarization, adding yet another complication to 
the spin-split 2DEG edge model.  Although the inclusion of the hyperfine 
interaction appears to just complicate an already complicated model, it 
might actually be useful as a tool for measuring the spatial electron 
spin variation.

As we have shown, the Overhauser shift can provide information about the 
local nuclear polarization, so it seems possible that the Knight shift 
can likewise be used as a probe of the spatially varying electron spin 
density near the edge.  At $\nu = 2$ the bulk of the 2DEG produces no 
Knight shift, since the net electron spin is zero.  Near the edge, 
however, there will be a region (the incompressible strip) of only one 
spin species, fringed by regions of unbalanced spin mixtures.  These 
regions of 2DEG would produce Knight shifts due to their net electron 
spin.  The summation of the Knight shifts from different regions of spin 
density should produce overstructure on the NMR absorption peaks.  Some 
of our data (not shown) show asymmetric NMR peaks with a slight bump on 
the left side, where a Knight-shifted peak would be expected to appear.  
Unfortunately, due to the switching noise of our sample, we were unable 
to accurately measure this overstructure, but we plan to pursue this 
method in the near future.  

\section{COMPARISON WITH OUR EARLIER EXPERIMENTS}

The experiments outlined in this paper are continuations of previous 
work by our group\cite{Wald} examining DNP effects using a similar 
experimental set-up\cite{setup}.  In this section, we review those 
previous results, noting that the observed hysteresis differed in 
important ways from the results reported in Section III.  We then 
discuss the origin of the differences between the two experiments.  We 
show that the voltages on the gates must be carefully chosen if they are 
to properly inject and detect the spin-polarized edge currents.  In the 
experiments of Ref. \cite{Wald}, this was not done, leading to what we 
now believe is an incorrect interpretation of the relative importance of 
the flip-flop and Zeeman terms in the experiments.  In particular, the 
hysteresis in Ref. \cite{Wald} was attributed entirely to the effects of 
flip-flop scattering, while we now feel that the influence of the 
nuclear Zeeman term was crucial to understand the experiments.   

In the experiments of Ref. \cite{Wald}, the $I-V$ curves displayed 
{\it symmetric} hysteresis. By this we mean that $|I|$  was greater when 
$V$ was being swept away from zero than it was when being swept toward 
zero, for both positive and negative $V$.  In other words, starting from 
the origin and sweeping $V$ from zero to (say) +1 mV to -1 mV to zero, 
the absolute current values were, in sequence: high, low, high, low.   
We explained this hysteresis by considering the currents carried by 
flip-flop scattered electrons.  Whenever the voltage changes sign, 
inter-edge scattering increases due to flip-flop scattering with the 
residual nuclear polarization, leading to an increased $|I|$.  We refer 
the reader to Ref. \cite{Wald} for a detailed explanation. 

In our more recent measurements (e.g. Fig. 3), the hysteresis was 
observed to be antisymmetric.   I is enhanced when sweeping $V$ away 
from zero for positive $V$ (because the spin gap is smaller due to the 
spin-up polarization), but suppressed for negative $V$ (because the spin 
gap is larger due to the spin-down polarization).  The hysteresis sweeps 
out a figure-8 (antisymmetric) rather than a pinched loop (symmetric).  
This asymmetric hysteresis is most naturally interpreted in terms of the 
nuclear Zeeman effect, as discussed in Section III.

Why is the hysteresis symmetry different?  The answer lies in the gate 
voltages applied to the QPCs that were used to inject polarized 
electrons into the scattering region. We observed antisymmetric 
hysteresis when we only partially depleted gates B and C, as shown in 
Fig. 1(a).  Upon increasing the voltage on these gates so that they 
became fully-depleted, the hysteresis became symmetric.  In the 
experiments of Ref. \cite{Wald}, fully depleted QPCs were used, 
resulting in symmetric hysteresis.

This observation led us to examine the AB split-gate by itself, in 
various states of depletion, to try to understand what was causing this 
hysteresis change.  Figure 6 shows the differential conductance through 
the AB split-gate as a function of $V$ for various values of $V_B$, with 
$V_A$ held at -1 V.  For $V_B >$  -0.35 V, the conductance is a fairly 
flat $e^2/h$, with some deviation at large negative V.  For more 
negative values of $V_B$, however, the conductance deviates drastically 
from $e^2/h$ for $|V| >$ 0.4 mV.  The value of the gate voltage $V_B$ at 
which this transition occurs is at the voltage at which the electron gas 
becomes fully depleted under the gate itself.  

Consider the paths of the edge channels near the AB split-gate, 
diagrammed in Figure 7.  The edge channels entering the split-gate from 
above are populated to the potential $\mu = -eV$ while the edge channels 
entering from the bottom are at zero potential.  If the AB split-gate 
forms a fully depleted QPC, the incoming and outgoing outer edge 
channels pass very close to each other while making their way between 
the gates, as shown in Figure 7(a).  If the bias $V$ is high, a large 
electric field will exist within the QPC, which could cause the 
electrostatic potential profile near the constriction to be deformed and 
cause unintended scattering and edge-state mixing (dotted lines).  For a 
partially depleted QPC, shown in 7(b), the edge states are very far 
apart, and little scattering is expected to occur.

We therefore conclude that that the electrons transmitted through a 
fully-depleted QPC (Fig. 7(a)) at high biases exhibit significant 
inter-channel scattering and thus are (a) not spin-polarized and (b) not 
populated up to the electrochemical potential m at which they entered 
the QPC.  On the other hand, for a partially depleted QPC (Fig 7(b)), 
the edge channels of different potentials remain macroscopically apart 
from each other, preserving the non-equilibrium current distribution 
even at large nonlinear biases.  As a result, the measurements and 
interpretaions reported in section III, using partially depleted QPCs, 
are more reliable than those given in Ref. \cite{Wald}, where fully 
depleted QPCs were employed.

Although we have shown that a full QPC displays complex behavior under 
high bias, the connection between this behavior and the change in the 
hysteresis loop remains poorly understood.  This is because the detailed 
behavior of the individual QPCs in this limit is not known; more 
experimental and theoretical work is required.  It should be possible to 
empirically measure the scattering matrix of such a QPC as a function of 
$V$ and the gate voltages, but we have not made an attempt to do so.  
Further, theoretical models of QPCs under high bias that takes into 
account the distortion of the electrostatic potential profile mentioned 
above should be developed.

\section{CONCLUSIONS}

We have observed $I-V$ asymmetry in scattering between spin-polarized 
edge states, and detected remarkably strong effects of GaAs nuclear 
spins upon these $I-V$ traces.  For forward bias, the $I-V$ trace 
displays a threshold which is nearly the bare Zeeman splitting, and for 
reverse bias the current increases only gradually with no apparent 
threshold.  We also observed hysteresis in these traces, which we 
interpret as being due to a combination of the dynamic nuclear 
polarization of the nearby nuclei and the hyperfine influence of the 
nuclear polarization on the electron energies.  The strength of the 
Overhauser field created by the polarized nuclei was found to be nearly 
as large as the external field itself.  The evidence for nuclear 
influence was supported by a series of NMR sweeps, which demonstrated 
that NMR absorption affected the current flow through the device.  From 
these experiments, we conclude that it is critical to consider the 
hyperfine interaction between Ga and As nuclei and the 2DEG in these 
systems, and that these interactions may be useful as a local probe of 
the edge.

We wish to thank Leo Kouwenhoven for useful discussions, and Bruce Kane 
and Jeff Beeman for technical assistance.  This work was supported by 
the Director, Office of Energy Research, Office of Basic Energy 
Sciences, Division of Materials Sciences, of the U.S. Department of 
Energy under Contract No. DE-AC03-76SF00098.  MRM acknowledges support 
from the NSF MRSEC for Technology Enabling Heterostructures grant 
DMR-9400415.

\newpage

\begin{figure}
\caption{(a) Schematic of device geometry for filling factor $\nu = 
2$.  Electrons of both spins enter from contact 1 at a bias $V$.  Only 
the spin-up edge channel is transmitted through gates A and B, and the 
electrons in this edge channel enter the scattering region where they 
can scatter into the grounded spin-down edge channel.  Scattered 
electrons then proceed to the current amplifier attached to contact 3 
(lower right) and are measured as current $I$.  Unscattered electrons 
disappear into the grounded contact 2 (upper right) and avoid detection.  
(b) AFM image of the device, with a 1 mm bar provided as a reference.  
The bottom gate was not used in these experiments, so it was grounded.}
\label{fig1}
\end{figure}

\begin{figure}
\caption{Landau level energy diagram near the edge of a 2DEG, for no 
bias (a), forward bias (b), and reverse bias (c).  The electron energies 
flatten out at the Fermi energy $E_F$ due to self-consistent 
electrostatic screening, forming compressible strips (flat regions, gray 
dots) and incompressible strips (sloped regions, black dots).  In 
forward bias (b), very little current flows unless $eV$ exceeds $g \mu_B 
B$, whereupon electrons can move readily from the inner to the outer 
edge channel.  In reverse bias (c), the current consists only of 
electrons that tunnel through the incompressible strip from the outer to 
the inner edge channel.}
\label{fig2}
\end{figure}

\begin{figure}
\caption{Spin diode $I-V$.  For forward bias, the current is small 
until $-eV$ reaches a threshold voltage comparable to the bare spin 
splitting $g \mu_B B$ = 0.175 meV.  In reverse bias, the current 
gradually increases with no apparent threshold.  The trace also displays 
hysteresis, with the $V$ sweep direction indicated by the arrows.  The 
two insets schematically show the flip-flop scattering between electron 
spins and nuclear spins for negative and positive bias.  The nuclear 
polarization is schematically shown for each step of the hysteresis 
loop, as discussed in the text.}
\label{fig3}
\end{figure}

\begin{figure}
\caption{$I-V$ traces taken at constant nuclear polarization.  For 
each trace, the nuclei were prepared by dwelling at a specified voltage 
$V_{dwell}$ for 60 seconds, then quickly changing the voltage to another 
value, measuring $I$, and returning to $V_{dwell}$ to maintain the 
polarization.  For $V_{dwell}$ = -1 mV, the nuclear polarization was up, 
and for $V_{dwell}$ = +1 mV, the polarization was down. The threshold 
voltage is shifted by the Overhauser effect of the prepared nuclear 
polarization on the electrons.}
\label{fig4}
\end{figure}

\begin{figure}
\caption{NMR absorption peaks, showing a marked change in current when 
the frequency of an in-plane AC magnetic field matches the splitting of 
a nuclear species (in this case, $^{75}$As).  The peaks shift linearly 
with $B$.  All plots were taken sweeping frequency from left to right.  
The $B$ = 7.0 T peak was swept at a much slower rate than the other two 
peaks, which have asymmetric lineshapes because the sweeping rate was 
comparable to the equilibration rate of nuclear repolarization.}
\label{fig5}
\end{figure}

\begin{figure}[b]
\caption{Plots of the differential conductance through the AB 
split-gate as a function of $V$ for various values of $V_B$.  The bulk 
filling factor $\nu = 2$, and $V_A$ = -1 V.  When gate B is only 
partially depleted (e.g. $V_B$ = -0.35 V), but still transmitting only 
one edge state, the conductance is basically flat at $e^2/h$, with a 
slight rise at nonlinear biases.  When gate B is depleted ($V_B <$ 
-0.35 V), the conductance deviates dramatically from $e^2/h$.}
\label{fig6}
\end{figure}

\begin{figure}[b]
\caption{Schematic of full and semi-QPCs, with the edge states flowing 
in the directions indicated by the arrows, and labeled by their 
electrochemical potentials.  In (a), both arms of the QPC are fully 
depleted.  The incoming and outgoing edge channels are forced to run 
close to each other inside the QPC, so if there is a large difference in 
their potentials, a large electric field exists within the QPC, which 
would distort the potential profile and cause unintended scattering and 
edge state mixing (dotted arrows).  In (b), gate B is partially 
depleted, but still only transmits one edge state, and the inner edge 
state is prevented from leaking through the region between the 
split-gates by a large $V_A$.  The incoming and outgoing edge channels 
are now far apart, preventing the scattering problems in (a).}
\label{fig7}
\end{figure}

\widetext

\begin{thebibliography}{10}
\bibitem{spinflip} The nuclear spin for all the relevant Ga and As 
isotopes is , so a "flop" process involves changing $I_z$ by 
$\pm\frac{1}{2}$ on its spin ladder.
\bibitem{setup} The measurement setup in Ref. 18 had the ground and the 
current amplifier reversed compared to our setup (Fig. 1a), so the 
measured current consisted of electrons which had not been scattered.
\end{thebibliography}
\end{document}